\documentclass[12pt]{iopart}
\usepackage{iopams}  
\usepackage{orcidlink}

\usepackage{graphicx}
\usepackage{xcolor}
\usepackage{subcaption}
\usepackage{times}
\usepackage{amsthm}
\usepackage{array}

\usepackage{longtable}
\usepackage{booktabs}

\newcommand{\ket}[1]{|#1\rangle}
\newcommand{\bra}[1]{\langle#1|}

\theoremstyle{plain}

\theoremstyle{definition}

\begin{document}

\title{Designing Minimalistic Variational Quantum Ansatz Inspired by Algorithmic Cooling}

\author{Soyoung Shin$^{1}$ \orcidlink{0009-0005-7449-196X}, Ha Eum Kim$^{2,3}$ \orcidlink{0009-0004-2614-0834}, Hyeonjun Yeo$^{1}$ \orcidlink{0009-0005-9368-306X}, Kabgyun Jeong$^{4,5}$ \orcidlink{0000-0001-7628-7835}, Wonho Jhe$^{1,*}$ \orcidlink{0000-0002-4716-5449}, Jaewan Kim$^{6,7,*}$ \orcidlink{0000-0002-1594-8924}}

\address{$^1$ Department of Physics and Astronomy, Seoul National University, Seoul 08826, Korea}
\address{$^2$ Department of Mathematics, Kyung Hee University, Seoul 02447, Korea}
\address{$^3$ Department of Physics, University of Illinois Urbana-Champaign, Illinois 61801, United State}
\address{$^4$ Research Institute of Mathematics, Seoul National University, Seoul 08826, Korea}
\address{$^5$ School of Computational Sciences, Korea Institute for Advanced Study, Seoul 02455, Korea}
\address{$^6$ Institute of Quantum Information Technology and Department of Physics, Yonsei University, Seoul 03722, Korea}
\address{$^7$ Quantum Universe Center, Korea Institute for Advanced Study, Seoul 02455, Korea}
\address{$^*$ Authors to whom any correspondence should be addressed.}
\ead{whjhe@snu.ac.kr and jaewan@kias.re.kr}

\date{\today}

\begin{abstract}
This study introduces a novel minimalistic variational quantum ansatz inspired by algorithmic cooling principles. The proposed Heat Exchange algorithmic cooling ansatz (HE ansatz) facilitates efficient population redistribution without requiring bath resets, simplifying implementation on noisy intermediate-scale quantum (NISQ) devices. The HE ansatz achieves superior approximation ratios with the complete network \textsc{Maxcut} optimization problem compared to the conventional Hardware efficient and QAOA ansatz. We also proposed a new variational algorithm that utilize HE ansatz to compute the ground state of impure dissipative-system variational quantum eigensolver (dVQE) which achieved a sub-$1\%$ error in ground-state energy calculations of the 1D Heisenberg chain with impurity and successfully simulates the edge effect of impure spin chain, highlighting its potential for applications in quantum many-body physics. These results underscore the compatibility of the ansatz with hardware-efficient implementations, offering a scalable approach for solving complex quantum problems in disordered and open quantum systems.

\end{abstract}

\maketitle

\tableofcontents

\section{Introduction}

Variational Quantum Eigensolvers (VQEs) have emerged as a versatile approach for estimating ground states and eigenvalues of quantum systems on near-term quantum hardware, often referred to as noisy intermediate-scale quantum (NISQ) devices \cite{Peruzzo2014, Preskill2018}.
Despite their promise, VQEs face significant challenges when applied to complex problems. One such challenge stems from the so-called barren plateau issue, where gradients vanish exponentially in large parameter spaces, making optimization infeasible for deep ansatz \cite{McClean2018, Cerezo2021}. Another challenge is the difficulty in accurately simulating systems that have complex local structures or impurities, which often leads to a high V-Score, indicating an increased number of parameters or layers needed to capture ground-state properties \cite{wu2024variational}. Such impurity systems, with their strongly inhomogeneous interactions and local disorder, frequently push VQE approaches to their limits in terms of both circuit depth and parameter optimization.

Recent advancements in quantum hardware performance and error mitigation techniques have paved the way for more sophisticated ansatz designs in noisy intermediate-scale quantum computers. As the industry continues to improve its quantum processors, researchers are developing a compact ans\"{a}tze that can leverage more quantum gates to overcome the barren plateau problem and enhance the  performance of variational quantum eigensolver.

IBM's latest Heron chip, capable of running 5,000 gates, demonstrates significant progress in hardware capabilities \cite{IBM2024} and QuEra's 10-logical qubit roadmap, promises to expand the possibilities for ansatz development further \cite{QUERA2024}. In response to these hardware improvements, many researchers have developed problem-specific ans\"{a}tze such as the Hamiltonian variational ansatz (HVA), quantum approximate optimization algorithm (QAOA), and unitary coupled cluster singles and doubles (UCCSD). 
The Hamiltonian variational ansatz methodologies have been introduced to alleviate some of these obstacles by systematically constructing circuits based on the problem Hamiltonian itself \cite{Wecker2015, Ho2019}. The HVA utilises system Hamiltonian to construct ansatz, which ensures that the variational evolution follows physically relevant pathways in Hilbert space. This structure can also be generalized to tackle non-linear dynamics and out-of-equilibrium processes, offering enhanced expressivity guided by principles from physics.

Such specialized ansatzs aim to target the spectral properties of difficult quantum systems more directly, thus helping mitigate barren plateau effects and reduce the circuit depth needed to capture long-range correlations or disorder-induced phenomena 
\cite{Ho2019, Wecker2015, Xanadu2024}. In parallel with these developments, quantum algorithmic cooling has been primarily investigated as a tool for state preparation \cite{Boykin2002}. By selectively transferring entropy from a target register to auxiliary qubits or an external bath, algorithmic cooling protocols can cool a subset of qubits into a purified state. While powerful, conventional algorithmic cooling implementations often require multiple qubit reset operations or a thermal bath, potentially complicating real-world applications on current hardware. Yet, the underlying concept of driving population redistribution through structured quantum operations has the potential to serve as a novel ansatz for variational algorithms.

This work proposes a quantum heat exchange algorithmic cooling protocol as an ansatz within the VQE framework by combining above ideas and progresses. By using unitaries designed to facilitate heat exchange among qubits by using auxilary qubits, the algorithmic cooling concept is harnessed to explore the state space more efficiently. Demonstrations show that this ansatz improves the performance on the NP-hard \textsc{Maxcut} problem, requiring fewer circuit layers to achieve higher approximation ratios compared to conventional VQE ansatz. Additionally, we proposed a new dissipative variational ansatz which proposed heat-exchange plays a critical role to compute the ground-state energy of quantum many-body systems with impurities. Notably, this approach circumvents the need for extensive Hamiltonian preprocessing or coupling to an external environment, thereby simplifying experimental requirements and make this algorithm suitable for the recent real quantum hardware. The synergy between algorithmic cooling-inspired layers and Hamiltonian-driven evolution suggests a robust and hardware-friendly strategy for tackling diverse classes of problems, including strongly disordered systems and other challenging contexts where standard VQEs may falter.

This paper is organized as follows.
In Section \ref{sec:HEansatz}, we delve into the fundamental principles and theoretical background of heat exchange quantum algorithmic cooling, and introduce the ansatz that will be utilized in subsequent investigations. Section \ref{sec:methods} describes the methodology for implementing the variational quantum eigensolver. In Section \ref{sec:case}, we analyze the proposed ansatz by assessing their performance on the NP-hard \textsc{Maxcut} optimization problem through both idealized simulations and executions on actual quantum hardware. We further apply these ansatz to a Heisenberg spin chain containing impurities, a prototypical quantum many-body system, to compute the ground state. Notably, we will demonstrate our calculations accurately capture the chains edge effects, thereby illustrating the effectiveness of the new algorithm. Finally, Section \ref{sec:conclusion} discusses the significance, limitations, and potential applications of the results.

\section{Heat-Exchange Algorithmic Cooling Ansatz} \label{sec:HEansatz}

\subsection{Variational Quantum Eigensolver}

The VQE algorithm is capable of determining the ground energy state and its corresponding quantum state.
This is based on the principle that the expectation value of a Hamiltonian $H$ with any arbitrary state $\ket{\psi}$ always exceeds the ground state energy $E_g$.
This statement can be expressed as follows:
\begin{eqnarray}
    E_g\leq\bra{\psi}H\ket{\psi},
    \label{eq:gr_leq_arb}
\end{eqnarray}
where equality holds when $\ket{\psi}$ is the ground state.
Given that the expectation value of the Hamiltonian is bounded below by $E_g$, VQE is designed to find a parameterized quantum state $\ket{\psi}$ that minimizes the expectation value of the Hamiltonian. This is achieved by using an appropriate ansatz for constructing $\ket{\psi}$ and an suitable optimization method. In details, the VQE algorithm consists of the following steps:
\begin{enumerate}
\item 
(\textit{Ansatz Preparation})
Based on the Eq. (\ref{eq:gr_leq_arb}), what VQE does is it tries to optimize the expectation value of the Hamiltonian. 
The algorithm starts by selecting a quantum circuit called an ansatz, which transform to $\ket{\psi}$. The ansatz is parametrized by a set of classical parameters, which can be adjusted to minimize $\bra{\psi}H\ket{\psi}$.

\item
(\textit{Quantum Simulation}) Each subterm of the Hamiltonian $H_i$, where the Hamiltonian is expressed as $H=\sum_i H_i$, is measured using the prepared state $\ket{\psi}$. By Collecting and summing up the expectation values of all subterms classically, the expectation value of the Hamiltonian is obtained.

\item
(\textit{Classical Optimization}) The calculated expectation value is sent to a classical computer, which adjusts the parameters of the ansatz to minimize the energy. This process is iteratively repeated until the energy fluctutation converges below a predefiend threshold.

\item
(\textit{Final Result}) The VQE algorithm returns the ground state energy and the optimized ansatz.
\end{enumerate}

In the VQE algorithm, the choice of ansatz is crucial, as it determines the algorithm's ability to explore the relevant Hilbert space and find the ground state. While highly expressive ansatz can represent a wide range of quantum states, their optimization often becomes computationally intractable. Hence, practical implementations prioritize trainability by selecting ansatz that balance expressibility and efficiency. For instance, the unitary coupled-cluster singles and doubles (UCCSD) ansatz performs well in chemical problems by narrowing the search space near the ground state, but is too resource-intensive for current quantum hardware due to its deep circuit and large gate count.

To overcome these challenges, physics-driven ansatz based on the symmetries and structure of the Hamiltonian have surfaced as a promising alternative.
By designing the ansatz with applying the properties of the system, such approaches enhanced the trainability and efficiency of VQE. For instance, Watanabe \emph{et al}. {\cite{Watanabe2023}} proposed a Hamiltonian-inspired ansatz tailored for strongly coupled electron systems, resulting in substantial improvements in optimization efficiency. These results highlight the potential of leveraging problem-specific insights to design ansatz that balance performance and hardware feasibility.

\subsection{Heat-Bath Algorithmic Cooling}

Quantum algorithmic cooling techniques aim to enhance the ground state population of target qubits by redistributing entropy within the system and, in some cases, with an external environment. Well-established traditional Heat-Bath Algorithmic Cooling (HBAC), is a technique that increases the polarization of qubits beyond thermal equilibrium by iteratively transferring entropy from the target qubits to a set of auxiliary bath qubits, which are periodically reset to their thermal states via interaction with a heat bath \cite{park2016heat,alhambra2019heat}. These foundational works establish the theoretical framework for HBAC, demonstrating its potential to surpass Shannon's entropy bound and achieve polarizations unattainable through conventional cooling methods.

HBAC begins with entropy compression, where reversible logic gates are employed to redistribute the entropy within the system. These gates are designed to transfer entropy from the target qubits—those used for computation or measurement—to the bath qubits, which act as temporary repositories for the excess entropy. At the next step, resetting the bath qubits to thermal equilibrium by brought into contact to a thermal reservoir, effectively removing entropy from the system. After the bath reset, the system is in a state of lower overall entropy, with the target qubits partially cooled. However, to achieve even lower temperatures and higher purities, the entire process is repeated. Through multiple iterations, HBAC can achieve cooling beyond the limitations of simple contact with a thermal bath, providing a powerful tool for preparing high-purity quantum states.

In addition to HBAC, several other non-unitary operations for finding ground states have been proposed. These include the Gutzwiller ansatz \cite{gutzwiller1963effect} and the Jastrow-Marshall ansatz \cite{huse1988simple}. Within the framework of VQE, Cobos \emph{et al}. \cite{cobos2024noise} introduced the concept of dissipative VQE. This method employs a sequential application of non-unitary and unitary operations, utilizing auxiliary qubits. While the probabilistic nature of quantum measurements typically precludes deterministic non-unitary operations, dissipative VQE achieves deterministic outcomes through the implementation of feed-forward operations.

\begin{figure}[ht]
    \includegraphics[width=15cm, angle=0]{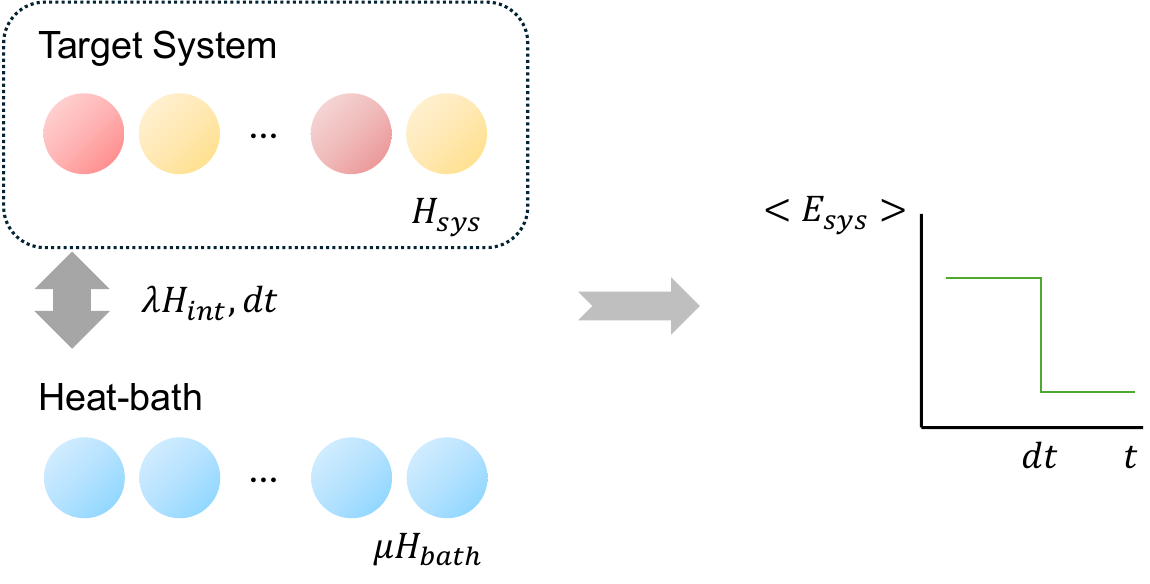}
    \caption{Heat exchange algorithmic cooling. The heat-exchange cooling method involves coherent interactions between target and bath qubits using the XX$+$YY Hamiltonian, without requiring bath qubit resets. At this process, the bath and target system interact one-by-one and the cooling efficiency mainly affected by the initial polarization of bath system.}
    \label{fig:HE}
\end{figure}

Insipred from these studies, we introduce heat exchange algorithmic cooling (HE) ansatz.
This ansatz leverages unitary operations to redistribute populations and coherences without the reset of a bosonic bath.
As depicted in Fig. \ref{fig:HE}, we consider a target system composed of copies of bosonic particles interacting with a bosonic heat bath, respectively. Without loss of generality, the interaction of each particle in target system and bath can be represented as
\begin{eqnarray}
\label{eq:HE_0}
    H_{\mathrm{int}}^{(j)}
    =J\left[a^\dagger_j b_j+b_j^\dagger a_j\right],
\end{eqnarray}
where $J$ is coupling strength, $a_j$ ($a^{\dagger}_j$) and $b_j$ ($b^{\dagger}_j$) are defined as annihilation (creation) operator of $j$th bosonic particle in target system and heat bath, respectively. Suppose that the heat bath is initialized as the ground state to be prepared to absorb the heat from the target system. Furthermore, the temperature is low enough that the number of particles of target system cannot exceed one. Then, Eq. \ref{eq:HE_0} can be represented as
\begin{eqnarray}
\label{eq:h_xx_yy}
    H_{\mathrm{int}}^{(j)}=
    J(\sigma^x_j\tau^x_j+\sigma^y_j\tau^y_j),
\end{eqnarray}
where $\sigma^k_j$ and $\tau^k_j$ are \emph{effective} Pauli matrices of ground and first excited $i$th bosonic particles of target and heat bath systems, respectively. 

This Hamiltonian represents the exchange of excitations between the two qubits, which is the basis for the cooling process. Unlike HBAC, which relies on a bath reset to eliminate entropy, the HE approach achieves cooling through a single evolution, and the advantages of the method are that it is more suitable for algorithmic implementation in quantum hardware in that it can cool quantum systems without any dissipative processes or external bath interactions.
This interaction causes transitions between the states $|10\rangle$ and $|01\rangle$ of the $j$-th target and heat-bath qubits:
\begin{eqnarray}
    e^{-iH^{(j)}_{\mathrm{int}}t}
    \ket{01}
    = \cos(Jt)\ket{01} -i \sin(Jt)\ket{10}\quad\mathrm{and} \\
    e^{-iH^{(j)}_{\mathrm{int}}t}
    |10\rangle 
    = i\sin(Jt)\ket{01} + \cos(Jt)\ket{10}.
\end{eqnarray}
At $Jt = \pi/4$, this results in a complete swap of populations between $|01\rangle$ and $|10\rangle$ enhancing the ground state population of $t_i$ if $b_i$ is more polarized (whole ground state), becomes equivalent to a partial iSWAP gate. 

As we introduced, heat exchange algorithmic cooling stands out for its simplicity, which relies solely on coherent unitary operations without coupling to an out-of-system heat bath during the cooling process. These features make it possible to algorithmically implement the target and bath as a closed system without exposing the quantum hardware to the external environment and facilitate the implementation of XX$+$YY interaction as a Pauli gate, making it easy to implement on real quantum hardware.

The achievable maximum ground-state population is essentially determined by the initial polarization of the bath qubit, making it easy to control the efficiency of cooling. Preparing the bath qubit to the ground state also makes it easy to achieve the upper limit of cooling efficiency for a single operation.

\subsection{Heat-Exchange Ansatz}

\begin{figure}[ht]
    \includegraphics[width=15cm, angle=0]{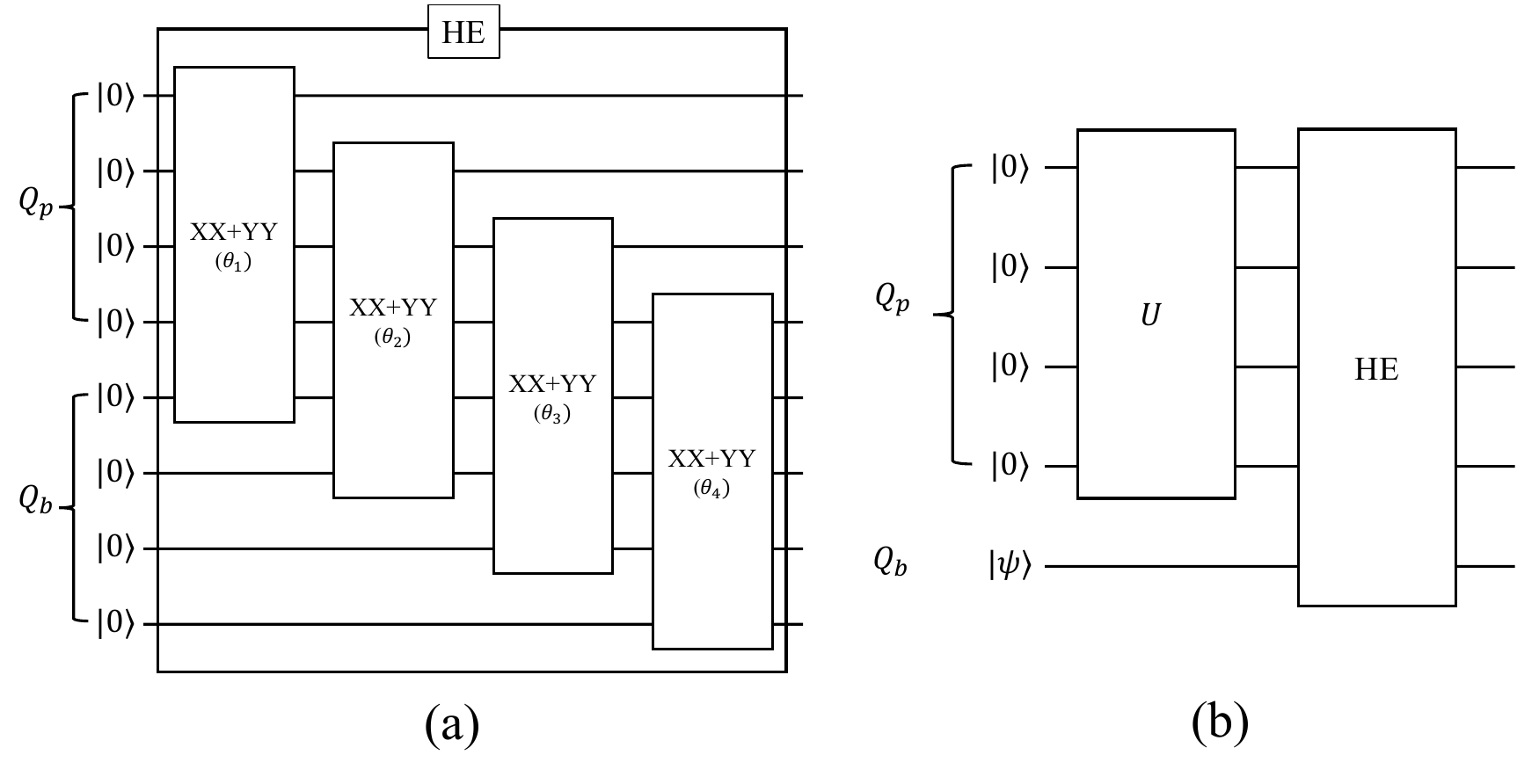}
    \caption{Proposed Ansatz. (a) Basic cooling block $C$ of 4 qubit system. $Q_p$ denotes the problem qubits which prepared at $|0\rangle$ states and $Q_b$ denotes the bath qubits prepared in $|1\rangle$ states. Problem qubits and bath qubits conneced by one-to-one XX$+$YY interactions. (b) Proposed dVQE \cite{Yoshioka2020} ansatz which dissipation block replaced by cooling block of (a). The circuit $U$ can be any ansatz compatible to the problem hamiltonian. Here we used \emph{hardware efficient} ansatz with  repeating structure of single-qubit and controlled X rotations.}
    \label{fig:ansatz}
\end{figure}

In this study, we propose two types of ansatz using heat exchange cooling as its key principle. First of all, for the quantum integer problems that can effectively calculate a solution with 1:1 interaction of heat exchange cooling, here we explore the ground state by ansats of Fig. \ref{fig:ansatz}(a) in the form of preparing the bath in a single qubit ground state and connecting it to the target system one-to-one. For problems in which the entanglement between the system qubits plays an important role, ansats in the Fig. \ref{fig:ansatz}(b) in which the dissipative term of the dVQE (i.e., dissipative VQE) ansats is replaced with cooling was used. 

Fig. \ref{fig:ansatz}(a) ansatz used to explore the solution of fully connected random weighted \textsc{Maxcut} problem at Section \ref{sec:methods}. For the Fig. \ref{fig:ansatz}(b) ansatz, it can be any ansatz compatible to the problem hamiltonian. Here we used proposed ansatz (b) to the 1D XXX Heisenberg chain of hamiltonian $\sum_i{(X_i X_{i+1}+Y_i Y_{i+1}+Z_i Z_{i+1})} + h\sum_i{Z_i}$, and we selected so called RealAmplitude ansatz, this is a hardware efficient ansatz with single R$_y$ and controlled X gate with proper repetition as a U block of Fig. \ref{fig:ansatz}(b). The details of new dissipative VQE algorithm will be covered by Section  \ref{sec:case}.

\section{Methods} \label{sec:methods}
\textbf{Variational Quantum Algorithm and Optimization Strategy}
\\
\\
In preparing actual VQE routine, one of the major challenges in optimizing quantum circuits include the presence of numerous local minima \cite{zhou2020quantum} and the noise introduced by quantum hardware. To avoid becoming trapped in a local minimum, surrogate models were proposed to identify good initial starting points before the advent of quantum algorithms \cite{powell1992theory, jones1998efficient}. To address noise in quantum circuits, several noise-robust optimizers have been developed, such as SPSA \cite{kandala2017hardware} and ImFil \cite{lavrijsen2020classical}. In this study, we combine a Gaussian process as a surrogate model with ImFil as a local optimizer to successively reach the global minimum, as proposed in Ref. \cite{muller2022accelerating}. We have found that this combination outperforms other optimization techniques, both with and without noise, which is consistent with the findings of Ref. \cite{muller2022accelerating}. 

For global search, the initial training number is limited to $0.1\times2^{N}$, where $N$ is the number of qubits and maximum Gaussian re-wetraining is 3 for all optimization process. For the local optimization process, the parameters were initialized by uniformly sampling random values within the bounds $[-\pi, \pi]$ for all rotational gates, and $[0, \pi]$ for the HE ansatz, considering its bit-flip nature. The maximum iteration number will be a control parameters to evaluate the performance of ansatz. 
\\
\\
\textbf{Setup}
\\
\\
All simulations and quantum circuit implementations were carried out using the Qiskit software development kit (SDK) \cite{qiskit2024}. Qiskit provides a comprehensive suite of tools for quantum computing, including modules for circuit construction, simulation, and execution on real quantum hardware. 
For ideal (noise-free) simulations, we utilized Qiskit Aer, a high-performance simulator capable of emulating quantum circuits with a high degree of accuracy. To evaluate the performance of our algorithm on real quantum hardware, we executed the quantum circuits on ibm strasbourg, a 127 qubits quantum processing unit (QPU) based on the Eagle r3 processor architecture, accessed via the IBM Quantum services. 
We applied the highest level of circuit transpilation optimization provided by Qiskit. The transpiler optimizes the quantum circuits by reducing gate counts, minimizing circuit depth, and mapping the circuits onto the hardware's topology while considering qubit connectivity and gate errors.
For error mitigation, we employed the Twirled Readout Error eXtinction (TREX) method \cite{trex2022}. 
The TREX is a measurement error mitigation technique that involves applying random Pauli operators (i.e., twirling) before measurement and then classically post-processing the results to reduce the impact of readout errors. By statistically averaging over different twirled circuits, TREX effectively suppresses biases in the measurement outcomes, enhancing the reliability of the experimental results. We perform 20,000
measurement shots for all gate-based algorithms.

\section{Case Study}
\label{sec:case}
\subsection{Combinatorial optimization problem}

Combinatorial optimization problems \cite{korte2011combinatorial} involve finding an optimal $N$-bit string that minimizes the cost function $C$, which is constructed based on possible combinations of bit strings. These problems have garnered significant attention as a potential area where quantum advantages are anticipated, due to their classical intractability and commercial value \cite{farhi2002quantum, zhou2020quantum, abbas2024challenges}. Among various combinatorial optimization problems, the \textsc{Maxcut} problem is a common target in this context \cite{Farhi2014QAOA, zhou2020quantum, wurtz2021maxcut}, due to its numerous connections with statistical physics \cite{barahona1988application} and machine learning \cite{boykov2001interactive}. 

Consider a graph $G = (V, E)$, where $V$ and $E$ represent the sets of vertices and edges, respectively. In this graph $G$, the \textsc{Maxcut} problem aims to find a partition that maximizes the number of edges cut by this partition. A key characteristic of the \textsc{Maxcut} problem is that it is NP-hard in the worst case, meaning there is no known polynomial-time algorithm to solve it. The cost function of the \textsc{Maxcut} problem can be expressed as follows: \begin{equation}\label{eq:costfn_maxcut} 
C = -\sum_{(i,j) \in E} \frac{1}{2}(1 - w_{ij} Z_i Z_j), \end{equation}
where $E$ denotes the set of edges in the graph, $w_{ij}$ represents the weight of the edge $(i, j) \in E$, and $Z_i \in \{1, -1\}$ denotes the partition of the $i$th vertex. The term ``weighted \textsc{Maxcut} problem'' is used to indicate whether the weight $w_{ij}$ differs across edges. Note that the weighted \textsc{Maxcut} problem is classified as NP-complete \cite{karp2010reducibility}.

In this study, we focus on the weighted \textsc{Maxcut} problem on a complete graph, where every pair of vertices is connected by an edge. This type of graph is frequently discussed in the literature \cite{inagaki2016coherent, kuete2023universal} because its well-defined structure allows researchers to consider only the randomness in edge weights, without the complexity introduced by variations in connectivity. The random weights used in this study is uniformly sampled between $0$ and $1$. Additionally, the weighted \textsc{Maxcut} problem on a complete graph encompasses the Sherrington-Kirkpatrick (SK) model of which weights are probability distributions with mean 0 and variance 1 \cite{harrigan2021quantum,farhi2022quantum}. The SK model \cite{panchenko2013sherrington} is strongly related to the spin glass problem, which is one of the fundamental issues in statistical mechanics.

\begin{figure}[ht]
    \includegraphics[width=15cm, angle=0]{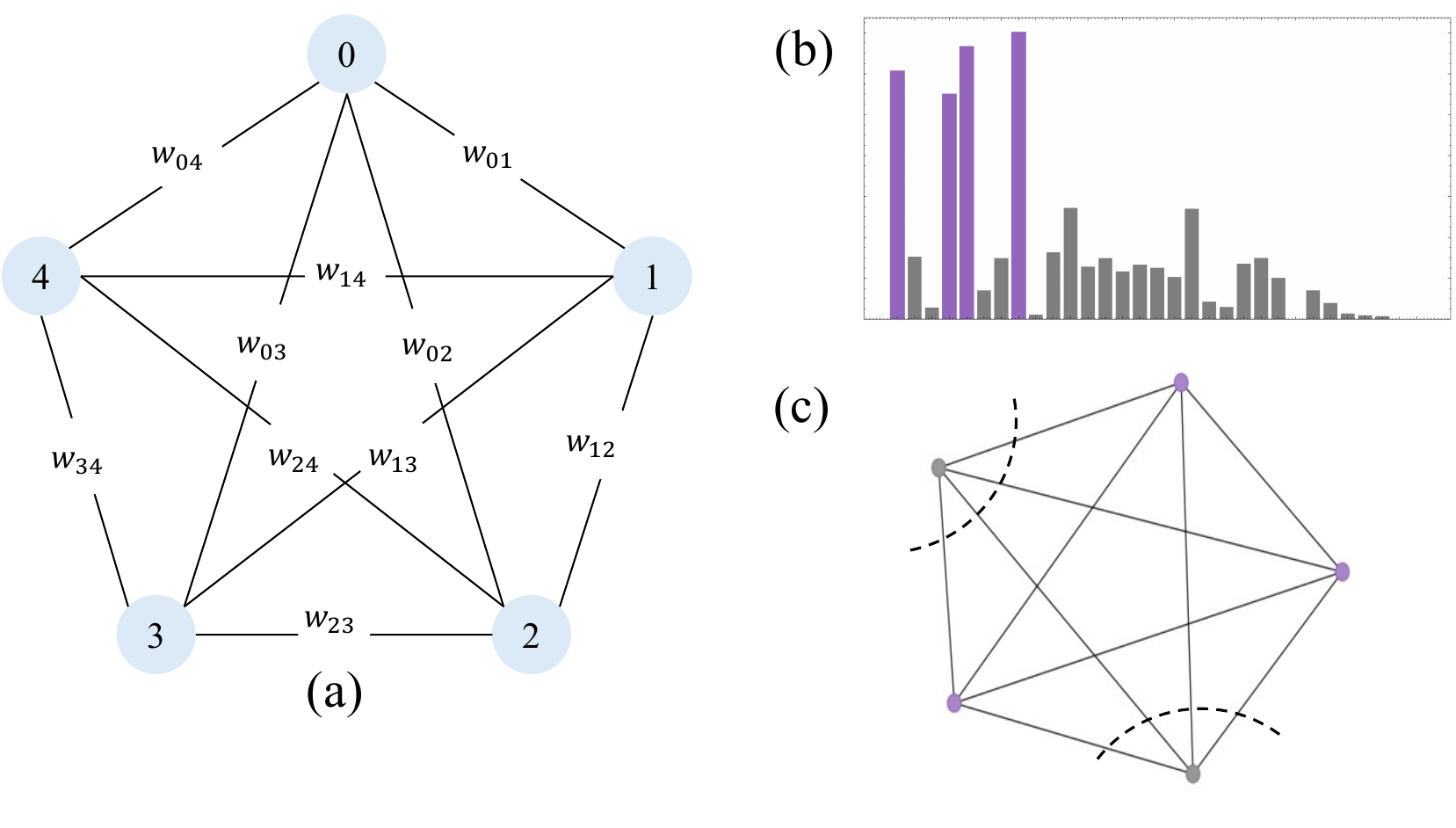}
    \caption{5 Node Complete graph weighted \textsc{Maxcut}. (a) shows the structure of the problem with random weights $w_{01},\ldots, w_{34}$. (b) shows typical histogram of circuit count of QAOA. The most probable configuration can be interpreted as a solution of \textsc{Maxcut} problem.}
    \label{fig:maxcut}
\end{figure}

In the numerical experiments, we adopt the following two metrics to evaluate the performance of the ansatz: approximation ratio and probability of finding a best cut. 
The approximation ratio \cite{Vazirani2001Approximation, Farhi2014QAOA, Hastings2019Classical, zhou2020quantum, Egger2021Warm} is one of the most widely used metric, denoted as $\alpha$, which measures how close the solution produced by an approximation algorithm is to the optimal solution. For the \textsc{Maxcut} problem, the approximation ratio is calculated as:

\begin{eqnarray}
    \alpha = \frac{C_{\mathrm{alg}}}{C_{\mathrm{opt}}} < 1,
    \label{eq:ar}
\end{eqnarray}
where $C_{\mathrm{alg}}$ is the cut value obtained by the algorithm and $C_{\mathrm{opt}}$ is the optimal (maximum) cut value. A higher $\alpha$ (i.e., closer to 1) indicates that the algorithm's solution is closer to the optimal cut. With a variational algorithm like QAOA and VQE and a given quantum state $|\psi(\theta)\rangle$ prepared by parameters $\theta$, the expected cut value $C_{\mathrm{alg}}$ is $\langle \psi(\theta)|H|\psi(\theta)\rangle$ with the problem Hamiltonian encoding the \textsc{Maxcut} objective function $H$. The other metric is the probability of finding the best cut after running the variational algorithms, which gives a direct intuition on how probable the ansatz returns an exact solution.

\begin{figure}[ht]
    \includegraphics[width=15cm, angle=0]{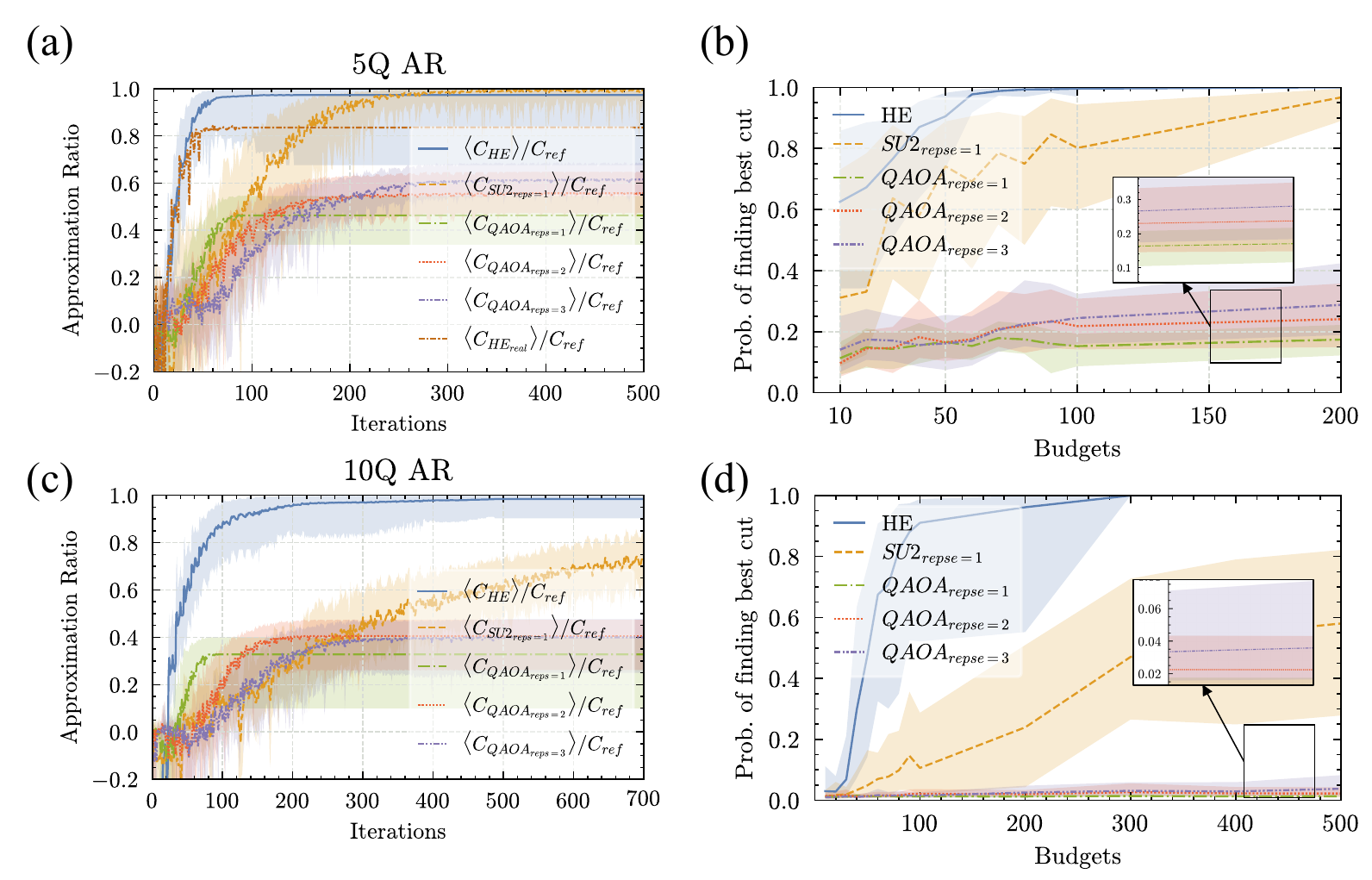}
    \caption{Numerical simulation and experimental results on approximation ratio and the probability of finding the best cut for various ansatz strategies applied to the \textsc{Maxcut} problem on random weighted complete networks of node sizes $n=5$  and $n=10$. (a, c) Approximation ratios over iterations show that HE consistently outperforms other methods, converging quickly to optimal values, while QAOA variants exhibit slower and lower convergence. $C_{HEreal}$ shows real hardware(ibm-strasboug) result of HE ansatz and it still overperform other ansatz. (b, d) Probability of finding the best cut as a function of budgets highlights HE's superior performance across all cases, with QAOA variants achieving significantly lower probabilities. Insets provide detailed views of specific budget ranges.}
    \label{fig:maxcut_result}
\end{figure} 

Figure \ref{fig:maxcut_result} shows the approximation ratio and the probability of finding the optimal cut for various ansatz strategies applied to the \textsc{Maxcut} problem on random weighted complete networks of node sizes $n=5$  and $n=10$. The simulated and experimental approximation ratio quantifies how closely the solution obtained by the algorithm approaches the optimal solution. The results indicate that newly proposed HE ansatz generally achieves a higher approximation ratio compared to QAOA and \emph{hardware efficient} ansatz reflecting its problem compatibility. Increasing the number of QAOA repetitions, $p$, leads to a notable improvement by align to previous researches in the approximation ratio. For instance, $p=2$ outperforms $p=1$, indicating that deeper circuits enable better exploration of the solution space. Notable feature is the performance of $p=3$ and $p=2$ becomes similar at $n=10$. This also well align to previous studies. The approximation ratio from real-device results (a) deviates slightly from the simulated outcomes, highlighting the influence of noise and hardware limitations in practical quantum experiments but still outperformed ideal simulation result of hardware efficient ansatz and QAOA anstaz.

\subsection{Quantum Many-body Problem with impurity}

The dissipative-system variational quantum eigensolver (dVQE) \cite{Yoshioka2020} is a quantum-classical hybrid algorithm designed to compute the non-equilibrium steady states (NESS) of open quantum many-body systems. It extends the variational quantum eigensolver framework, which traditionally focuses on closed quantum systems
The dVQE extends this methodology to open systems by introducing mechanisms to account for dissipation and environmental interactions.
At its core, dVQE reformulates the quantum master equation, particularly the Gorini-Kossakowski-Sudarshan-Lindblad (GKSL) equation, which governs the time evolution of a system's density matrix. The algorithm maps the mixed-state dynamics onto a vector representation in a Hilbert space with doubled qubits, leveraging the Choi isomorphism. This mapping enables the use of unitary quantum circuits to depict the dissipative dynamics. The Liouvillian operator, central to this reformulation, captures both the coherent (Hamiltonian-driven) and incoherent (dissipative) parts of the system's evolution. The dissipative contributions are modeled through jump operators that quantify energy loss or decoherence, ensuring the completely positive and trace-preserving nature of the evolution. The key innovation in dVQE lies in its cost function, derived from the squared Liouvillian operator $\hat{L}^\dagger \hat{L}$, ensuring that the algorithm identifies the steady-state solution where $\hat{L}|\rho_{SS}\rangle = 0$. This formulation allows the steady-state computation to be cast as a variational optimization problem. By optimizing the parameters of the quantum circuit, the algorithm minimizes the cost function, which corresponds to the NESS of the system under study.

Determining the ground state of lattice systems, especially those that are defective or have undergone a quench, is one of the major interests of condensed matter physics and materials science. A lattice represents a structured array of points (atoms, ions, or molecules) in space, and its ground state corresponds to the lowest energy configuration that the system can attain. Understanding the ground state of such systems is not only a theoretical pursuit, but implies many real-world applications, ranging from developing new materials to the study of fundamental quantum phenomena.
Considering this importance, recent work attempts to address this issue using VQE \cite{cobos2024noise, Buser2023, Liu2023, Zen2023,Tilly2022}, but making these methods compatible with current NISQ hardware, improving accuracy, and scaling up methods to cope with larger systems is also challenging.

In this section, we apply our method, i.e., HE dVQE ansatz in Fig.\ref{fig:ansatz}(b), to a Heisenberg chain with an external field to test the performance of our cooling method on a Hamiltonian that includes off-diagonal terms, in contrast to the diagonal Hamiltonian of Eq. (\ref{eq:costfn_maxcut}). The 1D Heisenberg Hamiltonian under consideration is as follows: \begin{equation} 
H = J\sum_{\langle i,j \rangle} (\sigma^x_i \sigma^x_j + \sigma^y_i \sigma^y_j + \sigma^z_i \sigma^z_j) + h\sum_i \sigma^z_i ,
\end{equation} 
where $\sigma^k_i$ are the Pauli $k$ operators for the $i$-th spin system, and $\langle i, j \rangle$ denotes nearest-neighbor spins in the 1D chain. This Hamiltonian has been  focused on numerous studies over several decades and remains a subject of ongoing interest \cite{giamarchi2003quantum, mikeska2004one, imambekov2012one}. Below the critical point $h_c = 2J$, this model belongs to the XY phase, which is equivalent to the Tomonaga-Luttinger liquid \cite{mikeska2004one}. Above the critical point $h_c$, the model transitions to a ferromagnetic phase, where spins are primarily aligned with the external field. For convenience, we set $J = 1$ and vary $h$ to explore both phases.

\begin{figure}[ht]
    \includegraphics[width=15cm, angle=0]{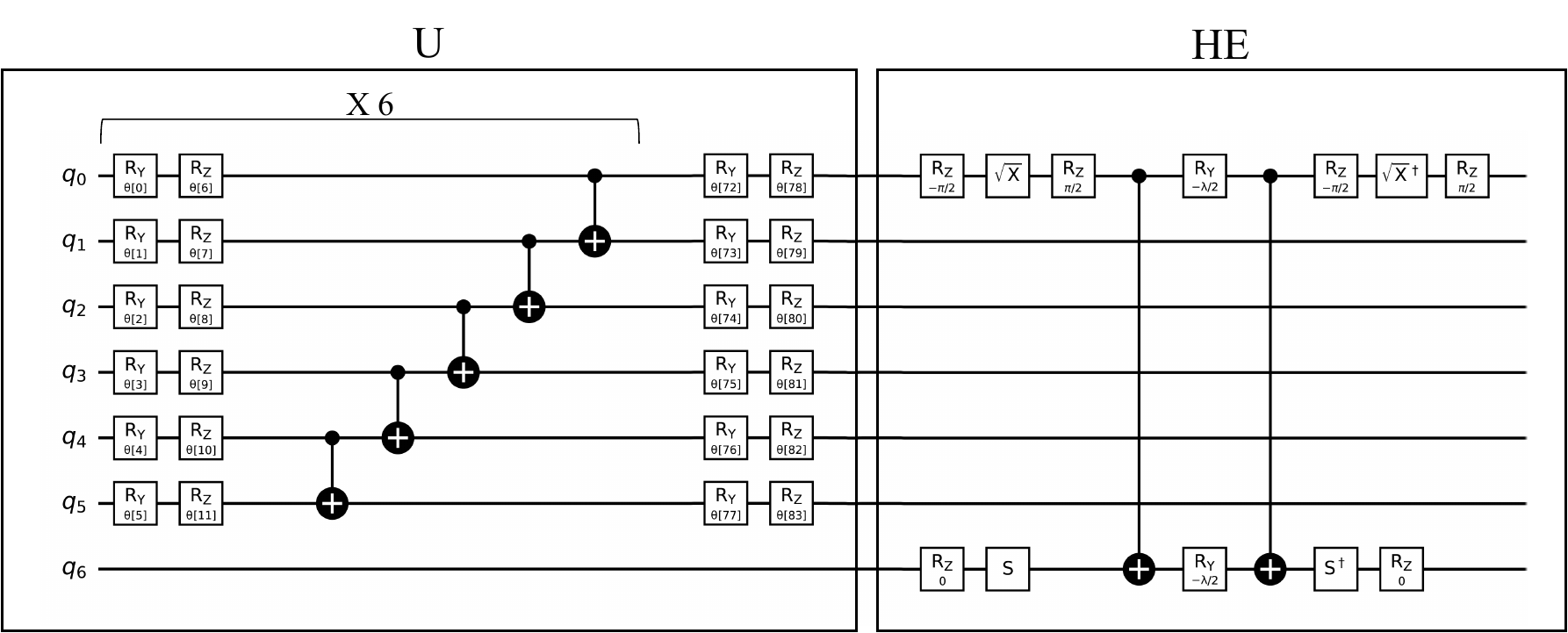}
    \caption{Ansatz circuit used to compute $N=6$ 1D XXX Heisenberg chain with impurity at 0-th qubit. The U block can be any ansatz. The HE block shows quantum circuit of parameterized HE ansatz works on 0-th qubit and bath ($q_6$). }
    \label{fig:dVQE}
\end{figure} 

We will study about the ground state energy and magnetization of this system with frozen state qubit which can be interpreted as impurity or quenched site. The system-bath interaction can be written 
\begin{eqnarray}
    2\lambda(\sigma_j^+\tau_j^- + \sigma_j^-\tau_j^+),
    \label{eq:int_xxyy}
\end{eqnarray}
where $\tau^k_j$ is Pauli $k$ operators for the $i$-th bath.
The raising and lowering operators $\sigma_j^\pm$ are defined as $\frac{1}{2}(\sigma^x_j \pm i \sigma^y_j)$, and same rule for $\tau_j^\pm$.
We consider the system's $i$-th qubit is dissipated due to the system-bath interaction defined as Eq. (\ref{eq:int_xxyy}).  
When interaction strength $g$ is small compared to the energy gaps of system, the time evolution of system's density matrix $\rho_s$ can be represented as Lindblad equation:

\begin{eqnarray}
    \frac{d}{dt}\rho_s(t) 
        &=& -i[H_s, \rho_s(t)] + \gamma(2\sigma_j^-\rho_s(t)S\sigma^+_j - \{{\sigma_j^+\sigma_j^-, \rho_s(t)}\}),
\end{eqnarray}
where $\gamma$ is decay rate proportional to $g^2$ and anti-commutation is defined as $\{A,B\}:=AB+BA$.
This master equation shows that the population of the excited state $|1\rangle$ at the 0-th qubit decays exponentially:
\begin{eqnarray}
    \frac{d}{dt}\langle S_0^+S_0^- \rangle = -2\gamma\langle S_0^+S_0^-\rangle.
\end{eqnarray}
This process cause the 0-th qubit's state settle into the $|0\rangle$, capturing the essence of energy dissipation. The off-diagonal elements involving the 0-th qubit can be computed by
\begin{eqnarray}
    \frac{d}{dt}\langle S_0^\pm \rangle = -\gamma\langle S_0^\pm\rangle
\end{eqnarray}
shows the decoherence effect, where the quantum superposition involving the 0-th qubit loses coherence over time.
By putting the dissipation and decoherence behavior of this interaction, neighboring qubits will experience modified dynamics due to the altered state of the frozen qubit, and the spin-spin correlation will be affected by this, impacting observable quantities like magnetization.

As mentioned by \cite{Yoshioka2020}, this dissipative VQE can capture the relaxation of the non-equilibrium state of a system by optimizing parameters of $U$ ansatz blocks with a dissipation. 

\begin{figure}[ht]
    \includegraphics[width=15cm, angle=0]{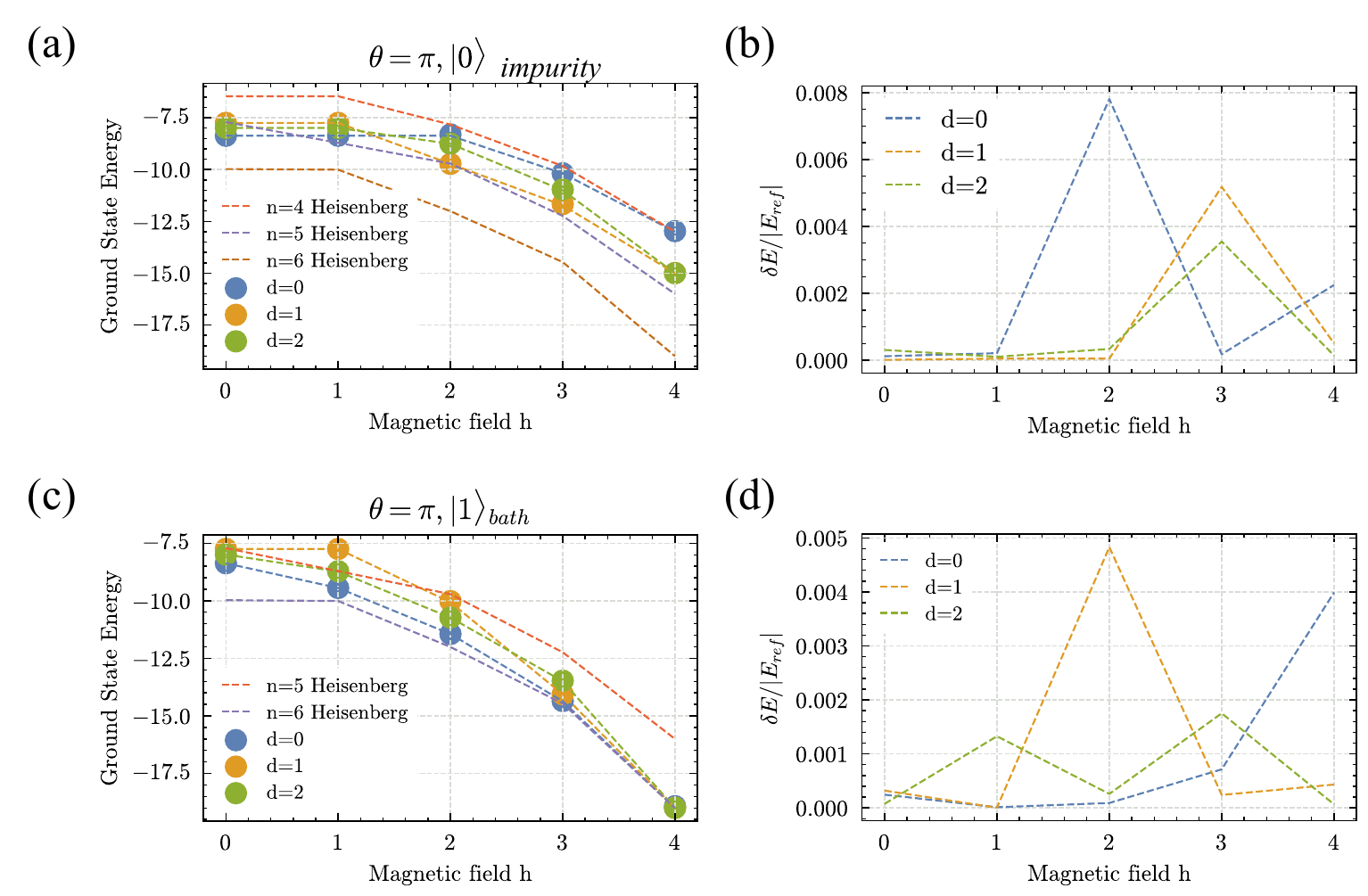}
    \caption{Computed ground state energy of XXX heisenberg rising chain with impurity. (a) and (c) shows the ground state energy of $n=6$ spin chain with impurities of distance $d=0,1,2$ from edge with magnetic field $h=0,1,2,3,4$. (b) and (d) shows the error of dVQE by comparing reference ground state energy. In both cases of impurity of $|0\rangle$ and $|1\rangle$ states, the maximum error rates are below $0.01$.}
    \label{fig:heisenberg}
\end{figure} 

Figure \ref{fig:heisenberg} shows the ground-state energy of a Heisenberg spin chain in the presence of impurities to $d=0,1,2$ from the edge when one impurity is present in a spin chain consisting of six spins, respectively, $\ket{0}$ or $\ket{1}$, and compare the ground-state energy calculated using dVQE with the reference value calculated using effective Hamiltonian numerically. In particular, we will look at the \emph{edge effect}, which is known to appear when impurities are at the edge.
Fig. \ref{fig:heisenberg}(a) show graphs of the ground state energy of the system for the magnitude of the external magnetic field for the case where impurities in the state of $|0\rangle$ are present at distances of $d=0,1,2$ from the edge. Except for the case where impurities are at the edge, the magnitude of the magnetic field exceeds $h_c=4$ have the same ground state energy.
The edge effect \cite{Hu1998, Qin1995} effectively reduces the length of the entire system when the impurity is present at the edge of the spin system with open boundary. Fig. \ref{fig:heisenberg}(a) shows that when the impurity is present at the edge, i.e., $d=0$, separated by the edge effect in the entire system if $h=4$ along with the adjacent spin, and showed that the Heisenberg spin chain with the ground state energy of the system equal the ground state energy of $n=4$ Heisenberg chain without impurity. For impurities present in the middle of the chain other than the edge, the effect of being isolated from the entire chain shows that the ground state is similar to that of the spin chain with $n=5$, confirming that dVQE captures the ground state energy and physical properties of the Heisenberg spin chain with impurities well.

\section{Conclusion} \label{sec:conclusion}
This study introduces a novel approach to addressing the challenges of ansatz design in quantum algorithms. It leverages a heat exchange algorithmic cooling (HE) ansatz and discusses the implications, advantages, and future potential of these contributions.
The proposed HE ansatz represents a leap forward in designing ansatz for open quantum systems. By using XX$+$YY interactions in the cooling mechanism, the ansatz achieves efficient redistribution of population and coherence without relying on bath resets or external dissipative processes. This approach simplifies implementation on NISQ devices, enhancing the method's practicality and scalability.
The performance of the HE ansatz is particularly evident in the dissipative VQE application, which combines the HE mechanism with optimization techniques to compute non-equilibrium steady states (NESS). The dVQE achieves a sub-$1\%$ error in estimating the ground state energy of contaminated Heisenberg-Ising chains, demonstrating its potential for studying quenched or disordered systems. These results open new avenues for exploring impurity-driven phenomena, with applications ranging from materials science to quantum chemistry.

Integrating the HE cooling block into dVQE extends its applicability beyond equilibrium systems to those involving dissipation and impurity effects. This ability to capture the dynamics of non-equilibrium states offers a robust framework for simulating complex physical systems, such as localized impurities in condensed matter or frozen-core approximations in quantum chemistry.
Moreover, the dVQE method's adaptability to real quantum hardware underscores its practical relevance. The algorithm's compatibility with the constraints of NISQ devices, including its tolerance to noise and low-depth circuit requirements, positions it as a powerful tool for near-term quantum simulations. The demonstrated scalability to larger systems and its high accuracy suggest its potential as a foundational method in quantum computation for disordered systems.

Future work could extend these methods to other classes of problems, such as quantum chemistry simulations and large-scale optimization tasks. Additionally, integrating the HE cooling mechanism with hybrid quantum-classical algorithms further enhances their utility in tackling real-world problems. 

\section*{Data availability statement}
The data generated and analyzed during the current study are available from the authors upon request.

\section*{Acknowledgments}
This work was supported by the National Research Foundation of Korea (NRF) through a grant funded by the Ministry of Science and ICT (NRF-2022M3H3A1098237) and partially supported by the Institute for Information \& Communications Technology Promotion (IITP) grant funded by the Korean government (MSIP) (No. 2019-0-00003; Research and Development of Core Technologies for Programming, Running, Implementing, and Validating of Fault-Tolerant Quantum Computing Systems). H.E.K. acknowledges support by Creation of the Quantum Information Science R\&D Ecosystem through the National Research Foundation of Korea funded by the Ministry of Science and ICT (NRF-2023R1A2C1005588). S.S. acknowledges support by ‘Quantum Information Science R\&D Ecosystem Creation’ through the National Research Foundation of Korea(NRF) funded by the Korean government (Ministry of Science and ICT(MSIT))(No. 2020M3H3A1110365).

\section*{Reference}

\bibliographystyle{alpha}

\bibliography{citation.bib}

\end{document}